# Ion-induced effects in GEM & GEM/MHSP gaseous photomultipliers for the UV and the visible spectral range

A. Breskin[*a], D. Mörmann[a], A. Lyashenko[a], R. Chechik[a]

F.D. Amaro[b], J.M. Maia[b,c], J.F.C.A. Veloso[b,d] and J.M.F. dos Santos[b]

[a] *Department of Particle Physics, Weizmann Institute of Science, 76100 Rehovot, Israel*

[b] *Department of Physics, University of Coimbra, 3004-516 Coimbra, Portugal*

[c] *Department of Physics, University of Beira Interior, 6201-001 Covilhã, Portugal*

[b] *Department of Physics, University of Aveiro, 3810-193 Aveiro, Portugal*

**Abstract**

We report on the progress in the study of cascaded GEM and GEM/MHSP gas avalanche photomultipliers operating at atmospheric pressure, with CsI and bialkali photocathodes. They have single-photon sensitivity, ns time resolution and good localization properties. We summarize operational aspects and results, with the highlight of a high-gain stable gated operation of a visible-light device. Of particular importance are the results of a recent ion-backflow reduction study in different cascaded multipliers, affecting the detector's stability and the photocathode's liftime. We report on the significant progress in ion-blocking and provide first results on bialkali-photocathode aging under gas multiplication.

## 1. Introduction

Most particle-physics, experiments having Ring Imaging Cherenkov (RICH) systems with gaseous photon imaging detectors, employ large-area CsI UV-sensitive photocathodes (PC) and wire-chamber electron multipliers [1]. In recent years there has been considerable progress in the development of other gaseous photomultipliers (GPMs) [2-7]. The R&D efforts have been generally motivated by the necessity to overcome some basic limitations of wire chambers. In these multipliers the avalanche develops at the wire vicinity, in an "open geometry", at a short distance (a few mm) from the PC. It results in significant photon- and ion-mediated secondary-avalanches formation, limiting the detector gain and its single-photon detection efficiency, and also affecting photon localization by broadening the charge induced on the readout elements. The ion-induced secondary-electron emission is particularly important in GPMs with visible-sensitive PCs, due to their low electron emission threshold [8]. Another important limitation of wire-chamber GPMs is the PC damage by avalanche-ion impact [9]; in wire chambers, like in parallel-plate avalanche chambers (PPAC) and in resistive-plate chambers (RPC), all avalanche ions are collected at the photocathode.

Our R&D efforts in recent years have therefore concentrated on the search for electron multipliers of a "closed geometry", with reduced photon- and ion-feedback probabilities. We have chosen cascaded "hole multipliers" of different structures, in which the avalanche that develops in successive multiplication stages is confined within the holes. We have investigated their physical properties and developed methods for reducing the avalanche-ion backflow (IBF), defined as the fraction of total avalanche-generated ions reaching the PC in a GPM; the methods discussed below are relevant to the stable operation of Time Projection Chambers (TPC), where the IBF relates to the fraction of ions reaching the drift volume. Recent results of ion-induced bialkali PC aging under gas-avalanche are presented.

---

[*] Corresponding author: Amos Breskin; Tel. 972-8-9342645; Fax. 972-8-9342611; amos.breskin@weizmann.ac.il

## 2. UV GPMs

The operation mechanism and properties of GPMs comprising cascaded Gas Electron Multipliers (GEM [10]) with semitransparent or reflective CsI UV-PCs, are summarized in [3,11]. Fig.1 shows a 4-GEM "reflective"-GPM with a PC deposited on top of the first GEM in the cascade [11]. It reaches gains $>10^6$, in a variety of gases, including $CF_4$ [12,13]. The resulting high sensitivity to single photons is due to the efficient GEM's optical screening, preventing avalanche-induced photons to hit the PC. The reflective-PC GPM has in addition very low sensitivity to charged-particles background, as discussed in [14]; relativistic-particle rejection factors >100 were recently demonstrated [15]. This property it is of prime importance in Cherenkov detectors operating under intense background, such as in relativistic heavy-ion experiments. The photoelectron detection efficiency, dictated by its extraction and transport into the holes, approaches unity at reasonable gains, as summarized in [3].

The compact structure of multi-GEM GPMs results in short multiplication times, yielding pulse-widths in the 10-20 ns range and single-photon time resolutions <2 ns [16]. The narrow avalanche width permits resolving close-by successive events; the width of the charge induced on the segmented readout anode can be tailored to cope with the readout scheme [17], e.g. by means of a resistive anode in front of the readout circuit [18]. 2D localization resolutions of the order of 100 μm RMS were measured with a 3-GEM detector coupled to a delay-line [17]. The IBF in cascaded GEM GPMs, reaching at best 10-20%, is discussed in paragraph 4.

The UV-sensitive "reflective" multi-GEM GPM is a mature technique; large-area detectors are under construction for a Hadron-Blind Cherenkov detector (HBD) of the PHENIX (RHIC-BNL) relativistic heavy-ion experiment [15]; others are developed for RICH [19]. Similar photon detectors, operated at cryogenic temperatures, are conceived for the XENON dark-matter experiment [20], for dark-matter and neutrino physics [21], for a Liquid-Xe PET project [22] etc.

With the goal of further reducing the IBF we have recently incorporated a Micro-Hole and Strip Plate (MHSP) [23] into a Multi-GEM cascaded multiplier [24]. The MHSP is a GEM-like hole-electrode with anode- and cathode-strips etched on its bottom face (Fig.2); the avalanche developed inside the hole is further multiplied on the anode strips. A significant part of the avalanche ions in the last strip-multiplication stage are collected at adjacent cathode strips and on the patterned readout cathode placed below the MHSP (Fig.2); this leads to a 4-5 fold smaller IBF in a single MHSP compared to that of a single GEM [24]. See paragraph 4 for details on the IBF in cascaded MHSP/GEM GPMs. Single-photon localization resolutions of ~100 μm RMS were reached in such detectors with a Wedge & Strip readout cathode [24].

Our latest development is the Thick GEM-like (THGEM) GPM [25]; it is manufactured by mechanically drilling sub-millimeter diameter holes in a thin printed circuit board followed by Cu-etching at the hole's rim. The operation and properties of these high-gain CsI-coated THGEM-GPMs are described in detail in these proceedings [26].

## 3. GPMs for the visible spectral range

Cascaded-GEM GPMs with visible-sensitive bialkali PCs have been extensively investigated; we shall point at the highlight of the results, while details are provided elsewhere [3, 8, 16]. The first significant outcome is the success in keeping semi-transparent bialkali PCs coupled to standard Kapton GEMs within indium-sealed detector packages for a month period, reaching QE values of the order of 13% at a wavelength of 435nm. However, the considerable ion-feedback at the PC limited the gain to values <1000 (spark limit); the secondary effects appearing well earlier [8]. A stable high-gain operation was reached by implementing an active ion-gating electrode [27]: a pulsed alternating-bias wire-plane, introduced within the cascaded-GEM structure suppressed the avalanche IBF to the PC by factor of ~$10^4$. This has brought about the next significant landmark: a gated 4-GEM bialkali-GPM (fig.3), permitted, for the first time, reaching gains of $10^6$ in the visible spectral range (Fig.4). The gating dead-time limits the counting-rate to ~0.1-1MHz; though it is acceptable in many applications, novel methods for ion-backflow suppression should permit DC-mode operation of visible-sensitive GPMs.

## 4. Ion backflow reduction

In view of the major role played by backflowing ions impinging at the PC, particularly in visible spectral range GPMs, considerable efforts were made to understand and to minimize this effect; past studies by others were carried out in multi-grid avalanche chambers [28], whereas current research focuses on cascaded multi-GEM [29,27,30] and multi-GEM/MHSP multipliers [31], and more recently, in other structures as summarized below. Practically, in our case, is the IBF defines the ratio between the ion charge collected at the photocathode and the electron charge at the last anode. Its dependence on the multiplier's geometry, gas and pressure, electric fields, etc. is generally well understood, as summarized in [32].

A key parameter controlling the IBF is the drift-field $E_{drift}$ above the first element in the cascaded multiplier (fig.1). It was shown that in detectors designed to detect charges deposited by gas ionization, the IBF can be kept low by applying low drift-field values; e.g., in TPCs $E_{drift}$ ~0.1 kV/cm. At such low fields the electron focusing into the GEM holes is very good; typical IBF values are between a fraction of a % to a few %, depending mostly on $E_{drift}$ and on the total gain [33,29,30,34], and partly on the geometry of the elements in the cascade [30]. In contrary, in GPMs the field at the PC surface must be high (above 0.5 kV/cm), to ensure an efficient photoelectron extraction into the gas [13], which contradicts the requirement for small IBF

Therefore, under such conditions, the IBF in multi-GEM GPMs could be reduced at best to levels of ~10-20% at a gain of $10^5$ [27,32] with a reflective PC deposited on the first GEM and with $E_{drift}$=0. Similarly, in a cascaded multi-GEM/MHSP with reflective PC, the IBF could be reduced to ~2-3% at effective gains of $10^5$-$10^6$ [31], due to the efficient ion collection by the cathode strips and plane (fig.2). However, this relatively small IBF value, adequate for CsI-GPMs, is not sufficient for eliminating the considerable ion-feedback observed in visible-sensitive GPMs with K-Cs-Sb PCs, of which the secondary electron emission probability $\gamma$, is 0.05-0.5 electrons/ion in $CH_4$ and $Ar/CH_4$ mixtures [35]. Such high $\gamma$ values require IBF smaller than $10^{-4}$ for stable operation at $10^5$ gains.

In an attempt to further reduce the IBF value, we have lately implemented a reversed-bias MHSP (R-MHSP) as a first multiplying element in the GEMs cascade [36]. In this particular MHSP operation mode [37], the positions of the anode and cathode strips are interchanged: the avalanche occurs only within the GEM-like holes, while the cathode strips attract a fraction of the up-flowing ions originating from avalanches in subsequent multiplying elements (Fig.5). However, as the anode strips attract a large fraction of the avalanche electrons, the ion trapping occurs at the cost of a drop in the number of electrons transferred from the R-MHSP to the subsequent element. In our first investigations [36] the IBF was reduced at best to ~1%, with two R-MHSPs preceding two GEMs at a total gain of ~$10^5$; this was obtained with a gain in the first R-MHSP <6. Due to the exponential nature of the single electron pulse-height spectrum many avalanches are of very small size and such a low gain in the first multiplying element does not provide full detection efficiency of the single electron events; such a low gain in the first R-MHSP would also affect the pulse-height resolution in multi-photon events and in ionization measurements in TPCs

Latest studies of multi R-MHSP/MHSP/GEM structures yielded better results [38]. Optimization of the potentials across the hole and between anode and cathode strips of the R-MHSP yielded gains >20 in this first element, ensuring full single-electron detection efficiency. Various detector configurations were investigated having a R-MHSP as a first multiplying element and a MHSP in the last multiplying stage, with GEMs or a second R-MHSP as additional intermediate elements. The drift field above the first element was kept constant at 0.5kV/cm. The best structures studied so far are a semitransparent PC coupled to a R-MHSP followed by 2 GEMs and a MHSP (Fig.6a), and 2 R-MHSPs followed by a MHSP, with a reflective PC on top of R-MHSP1 (Fig.6b); the IBF was suppresed down to 0.15% and 0.3%, respectively, at effective gains of $10^5$ (fig.7). These IBF values are by about two orders of magnitude smaller than that of cascades GEMs operated at similar gains and fields at the photocathode. The detailed results of this study are given in [38]; the search for more adequate ion blocking structures is in course.

## 5. Bialkali photocathode aging

The ageing of semitransparent K-Cs-Sb PCs under avalanche-ion impact is of great concern in visible-sensitive GPMs, and was recently investigated [39]. Bialkali PCs were produced on glass substrates, under vacuum of $\sim 10^{-10}$ Torr, as described in [8]. For each PC, the QE was measured upon production, in vacuum and then in the high-purity gas; the photocathode was coupled either to a 4-GEM multiplier or to a single-mesh electrode which formed a parallel-plate multiplier. The number of ions hitting the PC in each configuration was tuned by varying the multiplier's gain. The PC was illuminated with a focussed UV-LED light (375 nm) at a photon flux of $\sim 3*10^9$ photons/mm$^2$·s. QE measurements at two separate spots on the PC, one illuminated and subject to avalanche-ion flux and the other obscured, provided the ion-inuced aging - corrected for the decay by chemical processes with the gas impurities. Each ageing measurement lasted for the time required to accumulate 10-20 $\mu$C/mm$^2$ ion charge at the photocathode.. Some aging results of the bialkali PCs are presented in fig.8. One observes similar decay rate in all PCs investigated, up to an accumulated ion charge of $\sim$ 2 $\mu$C/mm$^2$, where the photocurrent (the QE) decays to ~80% of its initial value; then, the decay rates differ among the bialkali PC specimens, as a result of their different surface compositions (and in their initial QE) [40]. The decay of the bialkali PCs is similar to that of semitransparent alkali-halide PCs [9]. A detailed description of the method and discussion of the results is given elsewhere [39].

## 6. Summary

A novel generation of detectors with cascaded gaseous "hole-multipliers" may provide an improved solution for photon imaging. These fast devices, sensitive over a broad spectral range, have numerous applications including the imaging of Cherenkov light. One of their most important properties is the considerable reduction of the flow of avalanche-induced ions; with cascaded R-MHSP/GEM/MHSP multipliers the IBF to the photocathode was reduced by three orders of magnitude as compared to that of a MWPC-GPM. The back-flowing ion yield is proportional to the gain; therefore, with modern low-noise electronics and gains of $10^4$, we may expect the novel cascaded multipliers to limit the number of ions hitting the photocathode down to a few-to-a few tens per photon; the PC's lifetime will be ~1000 times longer as compared to that operated within a MWPC GPM. We can estimate the IBF effect on the bialkali PC aging: the 20% aging observed at an accumulated ion charge of 2 $\mu$C/mm$^2$, will occur after ~13 years of operation under a gain of $\sim 10^5$ at a photon flux of 1kHz/mm$^2$ (@ QE=30%).

Reduced ion-induced effects could be crucial for applications requiring the detection of large photon flux; it is mandatory for the DC-mode operation of visible-sensitive gaseous photomultipliers, prone to ion-feedback effects. Minimal IBF yields are also very relevant to TPCs, and with small drift fields, some of the methods discussed in this work have the potential of reducing the IBF to one ion per ionizing electron. An intensive search for further better electron-multiplier structures is in course.

This work was supported by the Israel Science Foundation, project 151/01 and by FCT (Lisbon) & FEDER, project POCTI/FNU/50218/03. A. Breskin is the W. P. Reuther Professor of Research in the Peaceful use of Atomic Energy.

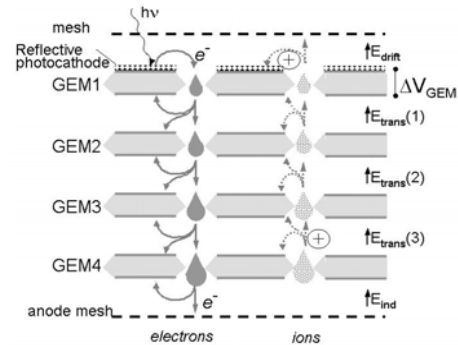

*Fig.1. A 4-GEM "reflective" GPM. Shown are the photoelectron trajectories, avalanche development and ion backflow in opposite direction.*

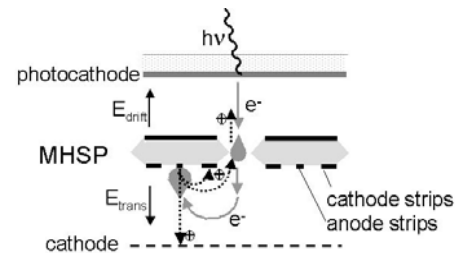

*Fig.2. The operation principle of a MHSP. Photoelectrons undergo "hole" and "strip" multiplications; a major fraction of the avalanche ions are collected by the cathode strips and plane.*

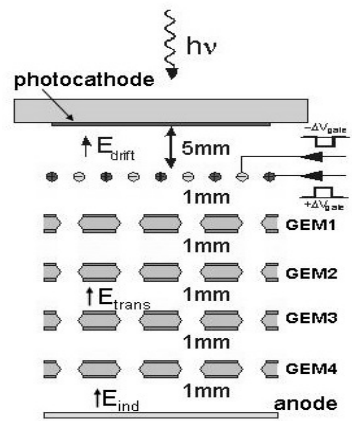

*Fig.3. A 4-GEM GPM with a semitransparent bialkali photocathode and an ion-blocking gate electrode.*

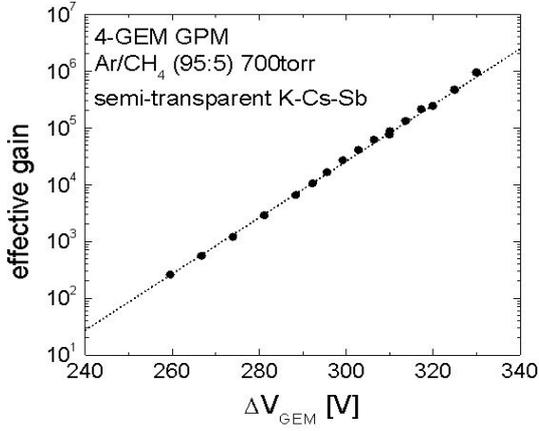

*Fig.4. Amplification curve of the gated visible-sensitive GPM.*

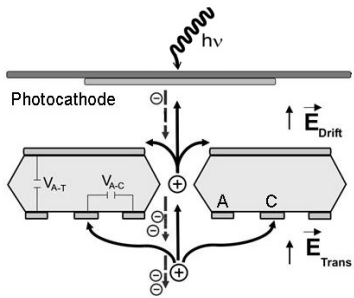

*Fig.5. The operation principle of a R-MHSP multiplier. Photoelectrons undergo "hole" multiplication; part of the electrons are transmitted to the successive element, while a fraction of the back-flowing ions from the latter is collected by the cathode strips.*

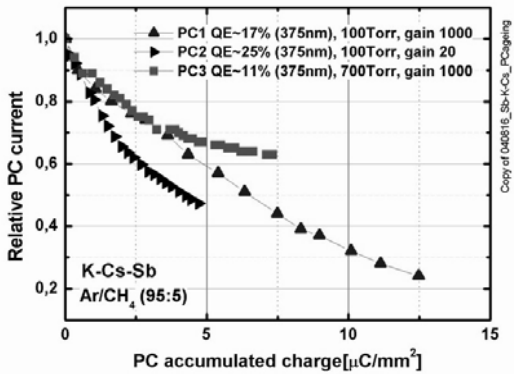

*Fig.8 Ion-induced K-Cs-Sb photocathode decay in $Ar/CH_4$ (95:5); initial QE values and pressures are indicated.*

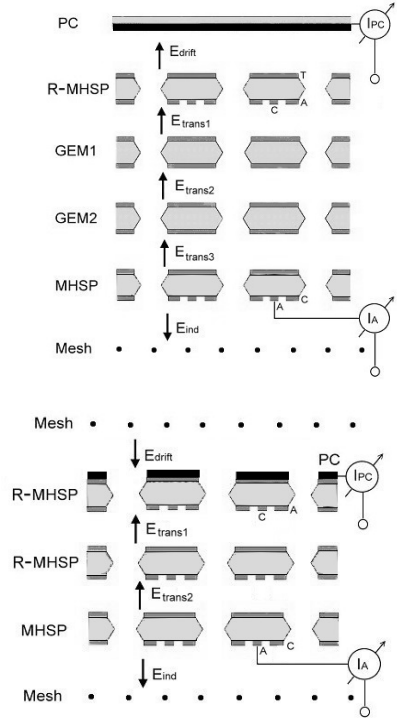

*Fig.6. GPMs with reduced IBF. **Top**: a R-MHSP coupled to a semitransparent CsI photocathode and followed by two GEMs and a MHSP; **bottom**: a reflective CsI photocathode on a R-MHSP, followed by a second R-MHSP and a MHSP.*

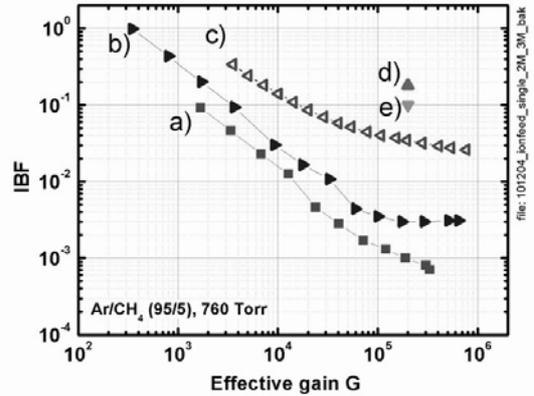

*Fig.7. The IBF values as function of total gain recorded in a) the "semitransparent"-GPM of fig.6a ($E_{drift}=0.5kV/cm$); b) the "reflective" GPM of fig.6b ($E_{drift}=0$); c) a "reflective" GPM with a triple-GEM followed by a MHSP ($E_{drift}=0$); d) a "reflective four-GEMs GPM ($E_{drift}=0$; low induction field); e) a "reflective" four-GEMs GPM ($E_{drift}=0$; multiplying induction field).*